\documentclass[11pt,onecolumn,amssymb,nofootinbib]{revtex4}
\usepackage{amsmath, amsthm, amscd, amssymb}
\usepackage{bm}
\usepackage{bbm}

\begin{document}

\title{\bf Planet-bound dark matter and the internal heat of Uranus, Neptune,
and hot-Jupiter exoplanets}
\author{Stephen L. Adler}
\email{adler@ias.edu} \affiliation{Institute for Advanced Study,
Einstein Drive, Princeton, NJ 08540, USA.}

\begin{abstract}
We suggest that  accretion of planet-bound  dark matter by the Jovian planets, and
by hot-Jupiter exoplanets,  could be a significant source of their internal heat.  The
anomalously low internal heat of Uranus would  then  be explained if the collision
believed to have tilted the axis of Uranus also knocked it free of most of its
associated dark matter cloud.  Our considerations focus on the efficient capture
of non-self-annihilating dark matter, but could also apply to self-annihilating
dark matter, provided the capture efficiency is small enough that the earth heat
balance constraint is obeyed.

\end{abstract}

\maketitle
The galactic halo dark matter mass density in the vicinity of the solar system is
currently believed to be about $0.3 ({\rm GeV}/c^2){\rm cm}^{-3}$, and corresponds
to dark matter that is gravitationally bound to the galactic center of mass, around
which it orbits along with our solar system.  Whether there is additional dark
matter in the solar system, either gravitationally bound to the sun, or to the
individual planets, is currently an open question.  Fr\`ere, Ling, and Vertongen
\cite{frere}
have pointed out that local dark matter concentrations in the galaxy may have
played a role in formation of the solar system, and this could give a rationale
for considering the possibility of both sun-bound and planet-bound dark matter.
This  suggestion is reinforced by recent simulations \cite{kuhlen} finding ``very
concentrated dark matter clumps surviving near the solar circle''; such clumps could be
natural nuclei for the formation of stars and planets.

Purely gravitational limits on the density of possible sun-bound or planet bound dark matter allow densities much larger than the galactic halo density.  Arguments based on planetary
orbits in Fr\`ere et al. \cite{frere} and the papers of Sereno and Jetzer \cite{jetzer}, Iorio \cite{iorio}, and  Khriplovich and Pitjeva \cite{khrip} place
a limit on the mass density of sun-bound dark matter of $\sim 10^5 ({\rm GeV}/c^2)
{\rm cm}^{-3}$. A comparison of lunar ranging and geodetic satellite tracking
observations \cite{adler} places a bound on the mass of earth-bound dark matter lying between the $\sim 384,000$ km radius of the moon's orbit and the 12,300 km radius of the LAGEOS satellite orbit of $4 \times 10^{-9}$ of the earth's mass; if such earth-bound dark matter were uniformly distributed, this translates into a mass density limit of
$\sim 6 \times 10^{10} ({\rm GeV}/c^2){\rm cm}^{-3}$.

Another source of limits on sun-bound and planet-bound dark matter
comes from considering the effect of dark matter accretion on solar
evolution  and earth and planetary heat flows.  Here assumptions
about the nature of dark matter non-gravitational interactions come
into play, and the assumption generally made is that dark matter is
self-annihilating.   From solar evolution, Fairbairn, Scott, and
Edsj\"o \cite{scott} find that stellar evolution  starts to be
altered by self-annihilating dark matter when the product of the
spin-dependent WIMP-nucleon cross section times mass density exceeds
$10^{-30}$ to $10^{-29}$ $({\rm GeV}/c^2) {\rm cm}^{-1}$.  From
considering the earth's heat flow, Mack, Beacom, and Bertone
\cite{mack} concluded that efficient capture in the earth of
self-annihilating dark matter with  the galactic halo density would
lead to a rate of energy deposition that exceeds the earth's
well-measured heat flow by a factor of about 100. This analysis, by
filling the gap between astrophysical constraints and underground
detector constraints, shows that galactic halo dark matter, under
the standard assumption that it is self-annihilating, cannot have
interaction cross sections with ordinary matter larger than the
usually assumed weak interaction cross sections.   Constraints on
galactic halo dark matter arising from considering annihilation in
Uranus were discussed by Mitra \cite{mitra}, and heating of Jovian
planets by galactic halo dark matter annihilation has also been
discussed in \cite{jovian}.  A possible role for galactic halo dark
matter in the heating of exoplanets was considered briefly in
\cite{mack}, again under the assumption that dark matter is
self-annihilating, but was dismissed as unlikely because of the
earth heat flow constraint.  Planetary heat production and volcanism
that may result from the  passage of the solar system through clumps
of galactic dark matter have been discussed in papers of Abbas,
Abbas, and Mohanty \cite{abbas}.

We wish in this note to reexamine the possible role of dark matter
in planetary heating, initially under the assumption that dark
matter is {\it not} self-annihilating, just as ordinary
baryonic/leptonic matter is not self-annihilating.  This could
happen, for example,  if dark matter is fermionic and consists of
fermions but not the corresponding antifermions.  It could also
happen if dark matter is bosonic and carries one sign of an additive
conserved quantum number, but not the opposite sign.
Non-self-annihilating dark matter would permit a large dark matter
interaction cross section with ordinary matter, making possible
efficient capture without violating the earth heat flow constraint.
Specifically, the analysis of the flyby anomaly in \cite{adler1}
shows that if the reported results are not an artifact, a dark
matter explanation would require dark matter masses well below a GeV
and a dark matter inelastic scattering cross section from ordinary
matter in the range between around $10^{-33}\, {\rm cm}^2$ and
$10^{-27}\,{\rm cm}^2$. Parameter values in this range are allowed
by existing constraints on dark matter masses and cross sections,
which are summarized in Sec. II of Mack, Beacom, and Bertone
\cite{mack}.  In Sec. IIA, these authors review the astrophysical
constraints, which require (for dark matter mass $m_d$  smaller than
a GeV) that the dark matter scattering cross section from ordinary
matter should be smaller than about $3 \times 10^{-25} (m_d c^2
/{\rm GeV}) {\rm cm}^2 $.   Direct detection constraints are
summarized in Sec. IIB and Fig. 1 of \cite{mack}, as well as in Fig.
3 of Gelmini \cite{gelmini}, and show that for dark matter masses
below a GeV, the entire cross section range between  $10^{-33}\,
{\rm cm}^2$ and $10^{-27}\,{\rm cm}^2$ is allowed.  For conventional
self-annihilating dark matter, this cross section range is almost
entirely excluded by the earth heat budget constraint, as shown in
Fig. 2 of \cite{mack}. However, for non-self-annihilating dark
matter, as noted by \cite{mack}, the earth heat budget constraint is
weakened by a factor of order $10^6$, and  parameter values of
interest for our present discussion are allowed.

The reason that the direct detection constraints reviewed in
\cite{mack} and \cite{gelmini} are not effective in placing limits
on dark matter masses much below 1 GeV,  is that these experiments
rely on detecting the recoil of a nucleon from which a dark matter
particle has scattered.  The smaller the mass of the incident dark
matter particle, the lower the kinetic energy of nucleon recoil, and
the harder it is to pick up this signature.  Hence experiments of
this type have a characteristic low mass cutoff in their sensitivity
to dark matter particles.  The same problem applies to the time of
flight beam dump experiment of Gallas et al. \cite{Gallas}, in which
one looks for events produced by particles that have detectable time
of flight differences from neutrinos, because for light, energetic,
dark matter particles, the time of flight difference that might
serve to distinguish them from neutrinos is not large enough. The
experiment of \cite{Gallas} has a lower mass limit of 0.5 GeV, and
for dark matter particles lighter than this places no constraints.
For dark matter particles in the mass range between 0.5 and 1 GeV,
interaction cross sections with nucleons between $10^{-29}$ and
$10^{-31} ~{\rm cm}^2$ are excluded if one assumes a production
cross times branching ratio  $\sigma \times {\rm br} =1000$ picobarn
per nucleon, whereas if one assumes $\sigma \times {\rm br}= 100$
picobarns per nucleon, there is no  excluded region for interaction
cross sections (see their Fig. 10).

The only type of accelerator search experiment that we have found
that does not have a low mass exclusion is the missing energy beam
dump experiment reported  by \AA kesson et al. \cite{Akesson}.  In
their Fig. 6, they use a theoretical model to extrapolate their
experimental results to give bounds on the production cross section
for stable neutral particles of masses 1 to 5 GeV.     For masses
below 1 GeV, their bound is in the range $1~{\rm to}~4 \times
10^{-31} {\rm cm}^2$.  However, such a production cross section does
not translate directly into an interaction cross section for dark
matter scattering on nucleons. For example, in QCD production of
particles by multiple gluon exchange, the phenomenological
Okubo-Zweig-IIzuka rule \cite{OZIrule} states that production
processes involving ``hairpin'' quark lines, in which the exiting
quark is not also an entering quark, are suppressed. Thus, in QCD
large classes of production processes are suppressed relative to the
cross sections expected from the corresponding scattering processes.
If analogous considerations apply to dark matter particles, then the
elastic scattering cross sections corresponding to the allowed range
of the experiment of \cite{Akesson} could be several orders of
magnitude larger, and would then encompass the whole range on which
we are focusing  our discussion here.  There are of course many
other accelerator experiments searching for new particles, but they
either assume that the new particles are unstable, and so decay
within a tracking device, or are charged, so that they leave tracks
themselves.  Finding neutral stable (or very long lived) particles,
such as putative dark matter particles,  is much more difficult,
which is why there are relatively few accelerator experiments
placing bounds.

To proceed, then, let us consider the collision of a dark matter
particle of mass $m_d$ and velocity $v_d$  with a medium containing
nucleons of mass $m_N$, and of sufficient optical depth that the
dark matter particle is certain to interact.  If the collision is
elastic,  a non-self-annihilating dark matter particle will multiply
scatter until it comes to rest, with an energy release in the medium
of $\frac {1}{2} m_d v_d^2$, which is smaller than the annihilation
energy $m_d c^2$ by the factor  \cite{beacom}
\begin{equation}\label{eq:elastic}
f_{\rm el}=  \frac{1}{2}\frac{v_d^2}{c^2}~~~.
\end{equation}
Consider next the case examined in \cite{adler1}, in which a dark matter primary particle of mass $m_d$ scatters inelastically on a nucleon into a secondary
particle of mass $m_d^{\prime}$, with $\delta m_d = m_d-m_d^{\prime}>0$, so that
the reaction is exothermic.  There are then two limiting cases.  If the secondary
scatters from nucleons strongly enough it will be trapped in the medium, and the
kinetic energy $\delta m_d c^2$ will be dissipated, giving an energy release which
is smaller than the annihilation energy $m_d c^2$ by the factor
\begin{equation}\label{eq:inel1}
f_{\rm inel\,1}=\frac{\delta m_d}{m_d}~~~.
\end{equation}
On the other hand, if the secondary scatters from nucleons only very weakly, so that
it escapes from the medium without energy loss, then the energy release is given
by the nucleon recoil energy $(1/2) m_N v_{\rm recoil}^2$.  As shown in \cite {adler1}, if $m_d^{\prime}$ and $\delta m$ are of similar order of magnitude, then
$v_{\rm recoil} \sim (m_d/m_N) c$, and so the nucleon recoil energy is
$(1/2) (m_d^2/m_N) c^2$, giving an energy release which is smaller than the annihilation energy $m_d c^2$ by the factor
\begin{equation}\label{eq:inel2}
f_{\rm inel\,2}=\frac{1}{2}\frac{m_d}{m_N}~~~.
\end{equation}
Clearly, other cases are possible, but we see already from the examples considered that
the factors $f_{\rm el}$, $f_{\rm inel\,1}$ and $f_{\rm inel \, 2}$ can all be
much smaller than unity.  For example, for a velocity $v_d$ in the range
$10 \,{\rm km}\,{\rm s}^{-1}$ to $50\, {\rm km}\,{\rm s}^{-1}$, characteristic of matter
orbitally bound to a solar system planet, $f_{\rm el}$ ranges from $5.6 \times 10^{-10}$ to $1.4 \times 10^{-8}$.  If $\delta m_d  << m_d$, then $f_{\rm inel\,1}$ is
very small, while if $m_d << m_N$, then $f_{\rm inel \,2}$ is very small.
So for non-self-annihilating dark matter, there are many possibilities for achieving
a much smaller energy release in the nucleon medium than the dark matter annihilation energy.

Consider now a planet with outward energy flow per unit area at its surface $H$.
Suppose that the planet is immersed in a dark matter cloud, with mass density $\rho_m$ and mean velocity
$v_d$ at the planet's surface. We will assume that the velocity $v_d$ is of the same order of
magnitude as the orbital velocity around the planetary surface $(GM_{\rm planet}/R_{\rm planet})^{1/2}$.  Continuing to denote by $f$ the
fraction of the dark matter annihilation energy that is deposited in the planet when  a dark matter particle is accreted, and including a solid angle factor of $1/2$, the condition for all of $H$ to be supplied by
dark matter capture is
\begin{equation}\label{eq:balance}
\frac{1}{2}\rho_m c^2 v_d f=H~~~,
\end{equation}
which gives for the dark matter density at energy flux equilibrium
\begin{equation}\label{eq:density}
\rho_m=\frac{1}{f} \frac{2H}{c^2 v_d} =\frac {K_{\rm planet}}{f}~~~,
\end{equation}
with
\begin{equation}\label{eq:kdef}
K_{\rm planet} = \frac{2H}{c^2v_d}\sim
 \frac{2H}{c^2} \big( \frac{R_{\rm planet} }{G M_{\rm planet} }\big)^{1/2}~~~.
\end{equation}
Using the planetary heat flow data given in de Pater and Lissauer \cite{depater}, we get the
following values for $K_{\rm planet}$ for Earth, Jupiter, Saturn, Uranus, and
Neptune,
\begin{align}\label{eq:kplanet}
K_{\rm Earth}=&  0.12({\rm GeV}/c^2) {\rm cm}^{-3} ~~~,\cr
K_{\rm Jupiter}=& 1.6 ({\rm GeV}/c^2) {\rm cm}^{-3} ~~~,\cr
K_{\rm Saturn} =&1.0  ({\rm GeV}/c^2) {\rm cm}^{-3} ~~~,\cr
K_{\rm Uranus} <&0.04 ({\rm GeV}/c^2) {\rm cm}^{-3} ~~~,\cr
K_{\rm Neptune}=&0.3({\rm GeV}/c^2) {\rm cm}^{-3} ~~~.\cr
\end{align}

As noted, these numbers have been computed using a dark matter velocity $v_d$ appropriate to planet-bound dark matter,
which is much smaller than the corresponding velocity associated with galactic halo dark matter.  As a check on our rather crude estimates, let us compare with the corresponding estimate of Mack, Beacom, and
Bertone \cite{mack} for the case of galactic halo dark matter. In their  ``maximum capture rate'' estimate, these authors
 take for $v_d$ the galactic halo dark matter average  velocity $270 \,{\rm km}\,{\rm s}^{-1}$, which is a factor of 34 larger than the earth surface orbital velocity of
$7.9 \,{\rm km}\,{\rm s}^{-1}$ used to compute the numbers in \eqref{eq:kplanet}.  Dividing the figure for the earth by 34 gives $K_{\rm Earth: ~halo~ dark~matter} =
0.0035 ({\rm GeV}/c^2) {\rm cm}^{-3}$, which in agreement with \cite{mack}, is two
orders of magnitude smaller than the estimated galactic halo dark matter density.
Hence, as concluded in \cite{mack}, for self-annihilating dark matter (corresponding to $f=1$),
accretion of galactic halo dark matter with perfect efficiency (corresponding to cross sections greater than $10^{-33} {\rm cm}^2$, for which the
optical depth of the earth is smaller than the earth's radius)  would give  too large
an internal energy generation for the earth, by two orders of magnitude.

However, let us now suppose that dark matter is not self-annihilating, so that this constraint on dark matter scattering
cross sections is no longer present,  and that
planets are typically  surrounded by a bound dark matter cloud.  The dark matter
mass density $K_{\rm planet}/f$ at the planetary surface that gives energy equilibrium is then, according to \eqref{eq:kplanet}, considerably larger than the
galactic halo density, and not outside the range determined by the gravitational
bounds on sun-bound and earth-bound dark matter.  So it then becomes reasonable to
hypothesize that some substantial fraction of the planetary internal energy generation comes from the  accretion of dark matter.  This fraction of the heat production coming from dark matter could account for unexplained residual heat production
in the earth \cite{mack}, the Jovian planets \cite{williams}, and in ``hot-Jupiter'' exoplanets \cite{exo}.  This proposal
assumes, and this is a topic for further study,  that
the surface depletion of the planet-bound dark matter cloud can be balanced by
accretion of planet-bound dark matter from the galactic halo dark matter, or from
dark matter bound to the sun or star around which the exoplanet orbits.

The hypothesis that planetary heat flows receive a significant contribution from
efficient accretion of planet-bound, non-self-annihilating dark matter, also can give a plausible explanation of the mystery of the anomalously low heat production from
Uranus.  Uranus and Neptune are structurally very similar \cite{wiki}, so one at first hand
would expect the internal heat flows to be similar.  However, in addition to the
difference in their heat flows, there is a second  well-known difference between Uranus and Neptune:  the axis of rotation
of Uranus is tilted 98 degrees with respect to the plane of the solar system, whereas the rotational axes of Neptune and the other Jovian planets have much smaller tilt angles ($<30$ degrees) with respect to this plane. The large axial tilt of Uranus is generally believed to be the result of a collision of Uranus
with a supermassive impactor.  Suppose now that the heat flux of Neptune and the other Jovian planets is primarily
associated with accretion from a planet bound-dark matter cloud.   Before its axis
was tilted by a collision, Uranus would also have been expected to have had an associated bound dark matter cloud, and a heat flux similar that of the other Jovian
planets.  But a collision at small impact parameter would have occurred within
the bulk of the Uranus-bound dark matter cloud, and plausibly could have knocked
Uranus out of the cloud, in analogy to what is observed in the colliding ``bullet''
galactic cluster merger  \cite{bullet}.  Once freed from its associated dark
matter cloud, Uranus would then be left with  a much lower internal heat production than Neptune and the other Jovian planets.

Finally, let us return to the case of self-annihilating dark matter, where the
energy release factor $f$ defined above is unity.  The suggestions we make concerning planetary heating could still apply if the dark matter interaction cross section
with ordinary matter is small enough so that the capture efficiency
is small, corresponding to parameter values below the heavily-shaded region in Fig. 2 of \cite{mack}.   Then the earth heat balance constraints of \cite{mack} can be
satisfied by galactic halo dark matter, but an excess of planet-bound dark matter
above the galactic halo density could lead to significant heating.  The formulas
\eqref{eq:balance}, \eqref{eq:density}, and \eqref{eq:kdef} would still apply, now
with $f<<1$ the capture efficiency rather than the ratio of the energy release to
the annihilation energy.  In the most general application of these formulas,
$f$ should be taken as the product of the capture efficiency times the ratio of the
energy release to the annihilation energy, since both of these factors can be
smaller than unity in the generic case.

I wish to thank Susan Gardner for an email in which she pointed out
the importance of the earth heat budget for my proposal in
\cite{adler1} of earth-bound dark matter, and in which she alerted
me to the  reference \cite{mack}, which led me to make the estimates
of $f$ in the inelastic case discussed above.   I also wish to thank
John Beacom for a stimulating email dealing with implications of
\cite{mack} for my paper \cite{adler1}, in which he made the
estimate of $f$ in the elastic case that I quoted above,  Ron Cowen
for a stimulating email and conversation about my initial posting on
planet-bound dark matter, and A. D. Dolgov and a referee for raising
the question of accelerator bounds on dark matter.   This work was
supported by the Department of Energy under grant no
DE-FG02-90ER40542, and I wish to acknowledge the hospitality of the
Aspen Center for Physics.

\end{document}